# Observation of the Meissner effect at room temperature in single-layer graphene brought into contact with alkanes

Yasushi Kawashima

Department of Precision Engineering, School of Engineering, Tokai University, Hiratsuka, Kanagawa 259-1292, Japan.

E-mail: kawasima@keyaki.cc.u-tokai.ac.jp

**There are claims of synthesis of a room temperature superconductor. However, these claims have not been officially accepted by scientific communities. Currently, the highest transition temperature (Tc) recognized in scientific articles is 135 K at 1 atm of Hg-Ba-Ca-Cu-O system which is a copper oxide superconductor. We packed graphite flakes into a ring-shaped polytetrafluoroethylene (PTFE) tube and further injected heptane or octane. Then we generated circulating current in this ring tube by electromagnetic induction and showed that this circulating current continues to flow continuously at room temperature for 50 days. This experiment suggests that bringing alkane into contact with graphite may result in a material with zero resistance at room temperature. In addition, we showed by means of AC resistance measurements using the two-terminal method that the resistances of graphite fibers brought into contact with various alkanes suddenly change at specific critical temperatures between 363 and 504 K. In this study, we show that after a magnetic field is applied to a single-layer graphene at room temperature, alkane is brought into contact with the single-layer graphene, then the graphene excludes the magnetic field immediately. This phenomenon demonstrates that the alkane-wetted single-layer graphene shows Meissner effect at room temperature. Furthermore, we applied a magnetic field perpendicularly to the annular single-layer graphene brought into contact with n-hexane and immediately removed the magnetic field. After that we observed that a constant magnetic field generates from this annular graphene for some time. In conclusion, the single-layer graphene brought into contact with alkane shows Meissner effect at room temperature, which provides definitive evidence for room temperature superconductivity.**

1．**Introduction**

The discovery of room temperature superconductors is one of the most important ultimate goals of researchers of superconductivity. If a room temperature superconductor is discovered, it would help solve the global energy problems and would lead to the development of new products.

In 1911, Onnes and his disciples discovered superconductivity of mercury at 4 K [1], and after that superconductivity of metal was discovered one after another, the highest $T_c$ of pure metal is 9.5 K of Nb [2]. Alloy superconductors were also discovered, the Tc of $MgB_2$ is 39 K, which is much higher than those of pure metals [3]. In 1986, copper oxides (cuprates) superconductors were discovered [4], which are the most important superconductors. After that, new cuprate superconductors were discovered one after another, the record of the highest $T_c$ was renewed by them, and the $T_c$ of cuprate ($YBa_2Cu_3O_{7-x}$) that was discovered in 1987 exceeded the nitrogen liquefaction temperature 77 K [5], which is much higher than the putative limit of $T_c$ (~40 K) predicted from the BSC theory [6]. Up to now, the cuprate superconductor $HgBa_2Ca_2Cu_3O_{8+x}$ has the highest critical temperature (Tc=135 K) at 1 atm [7].

To date, there are some claims of synthesis of a room temperature superconductor [8-12]. However, the scientific community does not officially accept these claims. It is generally accepted by scientists that the highest $T_c$ is equal to that (~135 K) of the cuprate superconductor $HgBa_2Ca_2Cu_3O_{8+x}$ [7].

In 2012, researchers in Germany reported data showing granular superconductivity at room temperature in water-treated graphite powder [11]. However, macroscopic superconducting currents and macroscopic Meissner effects are not clearly shown for this material.

To assert the room temperature superconductivity of a material, it will be usually necessary to present the following three independent phenomena as evidence. 1. The material has zero resistance at room temperature. 2. The material shows Meissner effect at room temperature. 3. These effects suddenly appear or disappear at a specific critical temperature above room temperature.

We inserted graphite flakes into a polytetrafluorethylene (PTFE) ring tube (shaped like a doughnut), injected octane or heptane into it, applied a magnetic field in the direction of the central axis of the ring tube and induced a ring current in the PTFE tube by rapidly turning off the magnetic field. We then showed that the ring current continues to generate a constant magnetic field for 50 days [12]. This phenomenon suggests that the current flowed through the ring with no resistance. Thus, we demonstrated that bringing alkane into contact with graphite may result in a material with zero resistance at room temperature. This is the first evidence of room temperature superconductivity mentioned above.

In addition, we measured the AC resistance of the PTFE tube packed with graphite fibers and alkane by using the two-terminal method while heating it. By this measurement, we observed the phase transition where the resistance sharply increased at a specific critical temperature and the rate of increase returned to its original value immediately [13]. This sharp change in resistance shows that the property of zero resistance of materials obtained by bringing alkane into contact with graphite suddenly appears or disappears at a specific critical temperature above room temperature. The observation of the sharp rise in resistance at a specific temperature above room temperature provides one of the evidence of room temperature superconductivity, that is, the third evidence mentioned above. If the rapid change in resistance observed is due to the phase transition from the superconducting state to the normal conducting state, the temperature at which the rapid resistance change occurred can be regarded as the $T_c$ of the conductor obtained by bringing alkane into contact with graphite. In the previous paper, we suggested that the $T_c$ of the superconductor obtained by bringing n-hexadecane into contact with graphite fibers exceeds 500 K [13].

The Meissner effect is a phenomenon in which a magnetic field-applied superconductor completely excludes the magnetic field from it during its transition to the superconducting state. Nowadays, the fundamental proof that superconductivity occurs in a material is the demonstration of the Meissner effect. Furthermore, the observation of the Meissner effect is the second evidence of the superconductivity mentioned above. If the Meissner effect was observed at room temperature in graphite brought into contact with alkane, then all three independent evidences of superconductivity would be obtained. If the conductor obtained by bringing alkane into contact with graphite is a superconductor, the superconducting current causing the Meissner effect to cancel the external magnetic field flows only at the contact portion between the graphite and the alkane. Therefore, the magnitude of the magnetic field that the alkane-wetted graphite can exclude will be very small.

In this study, we show that after a weak magnetic field is applied to a single-layer graphene, this magnetic field is immediately excluded by injecting alkane to the single-layer graphene. This shows that Meissner effect occurs in the single-layer graphene brought into contact with alkane. Furthermore, we applied a magnetic field perpendicular to the surface of annular single-layer graphene by a coil, and injected n-hexane to the surface. We show that the annular single-layer graphene generates a constant magnetic field for a while after turning off the coil power supply. This suggests that the current flows with no resistance on the annular single-layer graphene bought in contact with n-hexane at room temperature. Through these experiments, we demonstrate that a room temperature superconductor can be obtained by bringing alkane into contact with the single-layer graphene.

2．Methods

**2.1 Observation of the Meissner effect**

In the experiment of Meissner effect, a single-layer graphene formed on the copper foil was used as a sample. The sample was selected from the single-layer graphene deposited on a square copper foil (10 mm × 10 mm × 0.035 mm) purchased from Graphene Platform Corp. (Tokyo, Japan). As a device applying a magnetic field to the single-layer graphene, a PTFE cylinder embedded with a hand-made coil obtained by winding a 0.4 mm diameter enamel wire on a PTFE round bar with a diameter of 3 mm was used (see Fig. 1 (a)). As shown in Fig. 1 (a), a recess having a diameter of 17 mm and a depth of 1.5 mm was formed at the upper end of the cylinder. First, a PC permalloy (78.5wt%Ni, 4.3wt%Mo, 2.2wt%Cu, 0.47wt%Mn, Fe) circular foil (diameter 16 mm, thickness 0.012 mm) was placed on the recess of the end face of this cylinder (Fig.1 (a)). Next, the square single-layer graphene copper foil was placed on top of it using a wafer tweezers so that the graphene film surface was in contact with the permalloy foil. The single-layer graphene surface was placed on the permally foil facing downward. A top view of the PTFE cylinder when the single-layer graphene copper foil is placed is shown in the upper part of Fig.1 (a). As shown in Fig. 1 (a), after placing the graphene copper foil, the lower permalloy foil is not completely covered with the copper foil. An alkane was injected into a part of the permalloy foil which is not covered with this copper foil, using pipet. The injected alkane infiltrates the gap between the graphene surface of the copper foil and the permalloy foil to wet the surface of the single-layer graphene. To facilitate the infiltration of alkane, permalloy foil was manually deformed so that its central part was slightly dented. The reason the graphene film was not directed upward is as follows. When alkane is injected into the upward graphene film, the alkane evaporates and adheres to the tip of the Hall probe, which evaporates and takes heat away from the Hall element. As a result, the temperature of the Hall element decreases, so that a large error occurs in the measured value of the gauss meter.

A schematic diagram of the apparatus for observing the complete shield of the magnetic field due to the Meissner effect is shown in Fig.2. As shown in Fig. 2, the PTFE cylinder embedded with the coil was fixed to the bottom of a permalloy container (inner diameter 30 mm, depth 30 mm, thickness 1 mm) placed on an air core coil with a double-sided tape. After bringing the tip of the Hall probe into contact with the back side of the graphene copper foil, the probe was fixed by a fixing screw shown in Fig. 2.

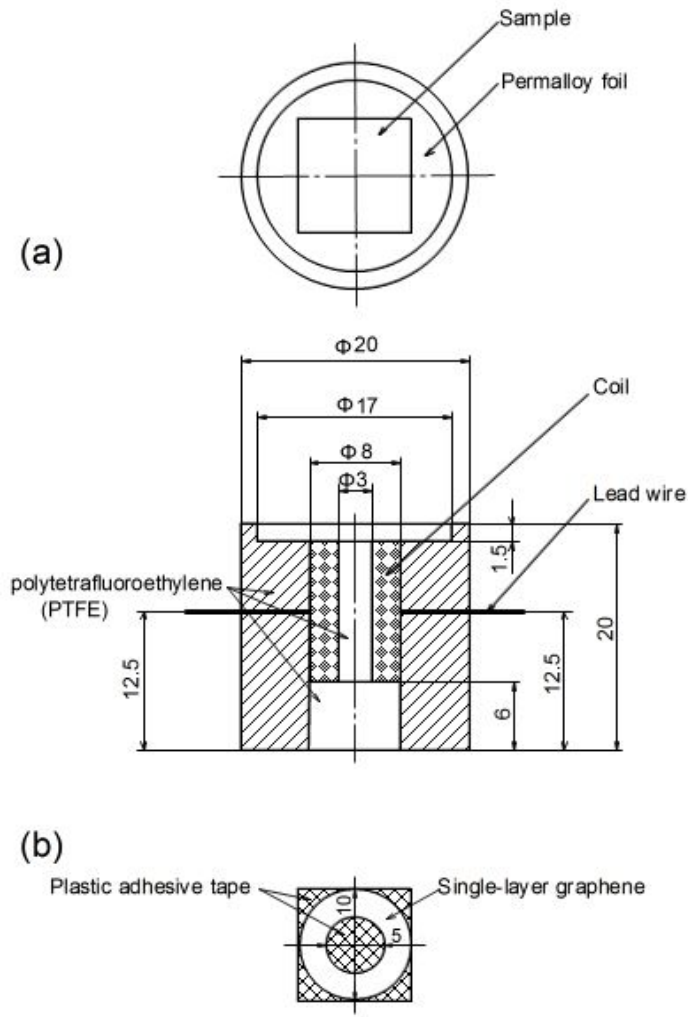

Figure 1  **A PTFE cylinder embedded with an electromagnetic coil for applying a magnetic field in a direction perpendicular to the single-layer graphene surface and an annular single-layer graphene sample.** (a)PTFE cylinder for applying a magnetic field to the sample. A recess of 17 mm in diameter and 1.5 mm in depth was machined at the upper end of the PTFE cylinder for placing a rectangular graphene sample. The upper part of the figure is a view of the PTFE cylinder seen from the top when a rectangular single-layer graphene sample is placed in this recess. (b)Annular single-layer graphene sample on $SiO_2$/Si.

Furthermore, the probe was raised by 1 mm by using the Z-axis translation stage (TBM-603, SIGMAKOKI CO., LTD.). Then the distance between the tip of the probe and the single-layer graphene surface becomes 1.035 mm. This probe was axial probe (HMNA-1904-VR, Lake Shore Cryotronics, Inc.), and the used Gauss meter was Model 455 DSP Gaussmeter (Lake Shore Cryotronics, Inc.). While watching the Gauss meter, the influence of geomagnetism was canceled by adjusting the current flowing through the air core coil under the permally container so that the axial component of magnetic field was nearly zero. Then, a magnetic field was generated in a direction perpendicular to the graphene surface by applying a current to the coil embedded in the PTFE cylinder by a DC current generator connected to the coil (see Fig. 2). About 30 seconds later, 8 µL of alkane (n-hexane, n-heptane) was injected into the permalloy foil. Changes in the magnetic flux density at approximately 1 mm from the graphene surface during the experimental process described above were stored every 0.2 seconds by a computer connected to a Gauss meter via a digital multimeter. The program was written in Excel VBA (Visual Basic for Applications).

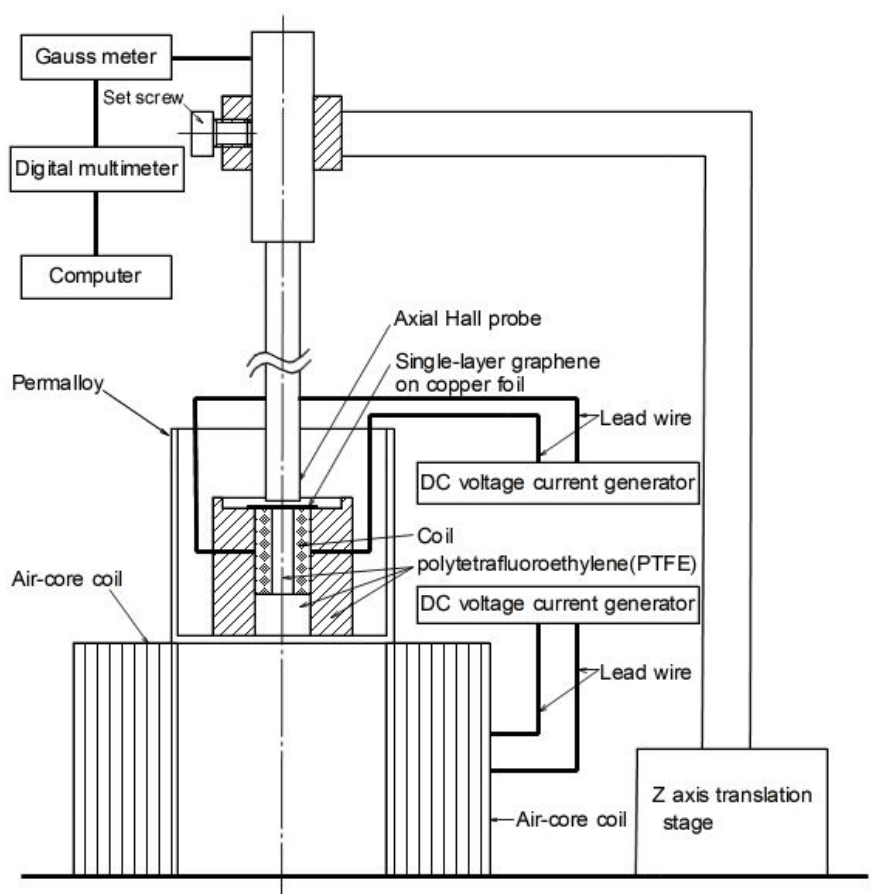

**Figure 2   A schematic view of an apparatus for observing Meissner effect occurring when an alkane is brought into contact with a single-layer graphene.**

## 2.2 Measurement of circulating current flowing at zero resistance in annular graphene brought in contact with n-hexane

A single-layer graphene placed on a silicon oxide film formed on p-type silicon was used as a sample. The sample used was selected from single-layer graphene deposited on $SiO_2$/Si (10 mm×10 mm×0.525 mm) purchased from Graphene Platform Corp. (Tokyo, Japan). This $SiO_2$ film is an oxide film with a thickness of 100 nm±10% formed by thermal oxidation on p-type silicon (electrical conductivity: 10 S/m). As shown in Fig. 1(b), a plastic tape having a thickness of 0.12 mm was attached to this sample to form an annular single-layer graphene. The annular graphene has an outer diameter of 10 mm and an inner diameter of 5 mm. The PTFE cylinder embedded with the coil shown in Fig. 1 (a) was placed in a permalloy container having an inner diameter of 30 mm, a depth of 70 mm and a thickness of 1 mm. The cylinder was fixed to the bottom of the container by a double-sided tape having a strong adhesive force. As in the case of the Meissner effect observation experiment, the permalloy foil was placed on the recess of the PTFE cylinder in which the coil was embedded. Next, the sample was placed on the permalloy foil such that the single-layer graphene film was in contact with the permalloy foil. The axial Hall probe was brought into contact with the back surface of the sample, and it was fixed with the screw. After that, the Hall probe was raised 1 mm by the Z axis translation stage. In this case, the distance between the graphene surface and the Hall element is about 1.5 mm. A coil current for generating a magnetic field of 0.15 G using the coil embedded in the PTFE cylinder was determined while watching the Gauss meter. The experiment was carried out as follows. 20 μL of hexane was injected onto the permalloy foil. After 30 s, the magnetic field of 0.15 G was applied in the direction perpendicular to the graphene surface by turning on the coil power supply. After holding for 30 s, the coil power supply was turned off. Subsequent changes in the magnetic field were recorded every 0.2 s by the computer connected to the Gauss meter.

## 3. Results and Discussion
## 3.1 Observation of the Meissner effect at room temperature

The measurements were made while the influence of the geomagnetism perpendicular to the single-layer graphene surface was canceled by the magnetic field generated by the air core coil set just under the permalloy vessel. Figure 3 shows the changes with time of the magnetic flux density measured by the Hall sensor at a position about 1 mm away from the single layer graphene surface. There are small up and down fluctuations in the measurement data, but these are due to the instability of the air core coil excitation power supply. The power source for exciting the coil embedded in the PTFE cylinder for applying a magnetic field perpendicular to the graphene surface was turned on, and after about 30 seconds, the alkane was injected into the permalloy foil. The changes with time of the magnetic flux density measured by the Hall element during this period is shown in Fig. 3.    Figures 3 (a) and 3 (b) show the changes with time in the magnetic flux density when hexane is injected, and Figs. 3 (c) and 3 (d) show the changes with time in the magnetic flux density when heptane is injected. Figures 3 (a) and (b) show that the measured magnetic flux density abruptly changes in the direction opposite to the direction of the applied magnetic field immediately after n-hexane injection to the single-layer graphene, and it converges to zero. Figures 3 (c) and (d) also show that the magnetic flux density measured by the Hall sensor abruptly changes in the direction opposite to the direction of the magnetic field applied immediately after heptane injection, and the magnetic flux density converged to almost zero. These results show that the single-layer graphene completely shields the magnetic field by bringing alkane into contact with the graphene, indicating that the graphene surface completely excludes the magnetic field from the graphene as soon as it touched the alkane. That is, the Meissner effect was shown to occur in the graphene brought into contact with alkane. To clearly observe the Meissner effect, it is necessary to completely cover the entire surface of the copper foil with graphene. From the viewpoint of reproducibility of the experiment, it may be desirable to use two-layer graphene which has been processed twice.

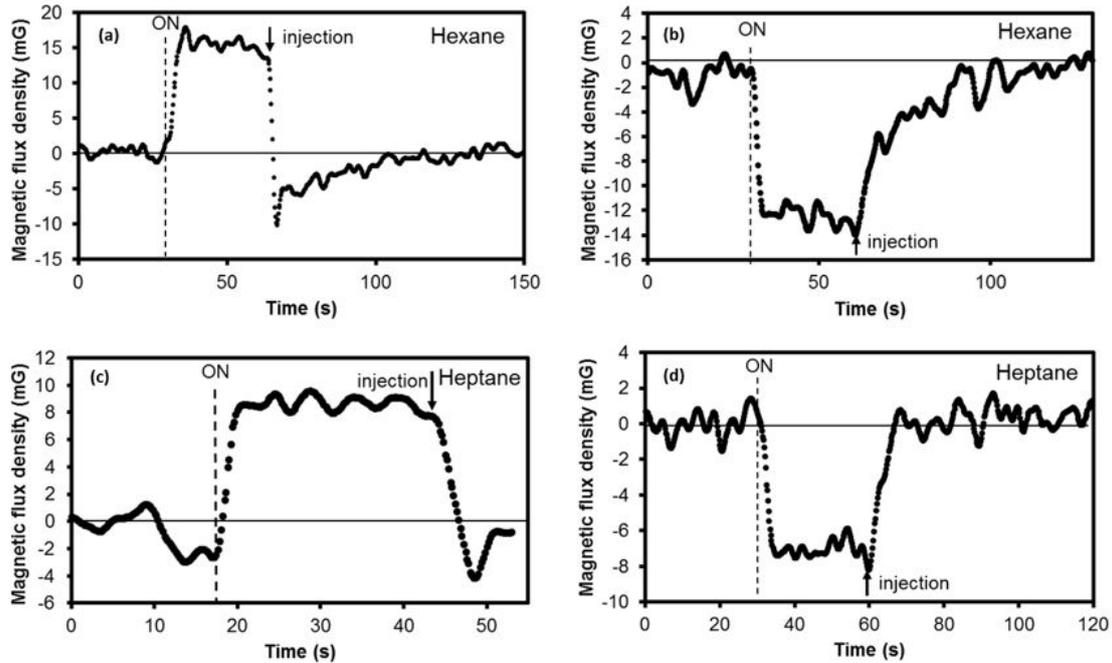

Figure 3    **Changes with time of magnetic flux density measured by the Hall element until the magnetic flux density becomes almost zero by injection of alkane.** Here, after turning on the power supply of the electromagnetic coil for applying a magnetic field in a direction perpendicular to the single-layer graphene surface, the alkane was injected about 30 seconds later. (a) and (b) are the changes with time of magnetic flux density in the case that n-hexane is injected, and (c) and (b) are those when injecting n-heptane. The direction of the magnetic field applied by the coil in the case of (a) and (c) is opposite to the case of (b) and (d). Arrows indicate the time of alkane injection.

### 3.2 Measurement of constant circular current flowing on annular single-layer graphene surface in contact with n-hexane at room temperature

Figure 4 shows the changes with time of the magnetic field in the direction perpendicular to the annular graphene surface at the position about 1.5 mm away from the annular graphene surface immersed in hexane after cutting off the current flowing through the coil embedded in the Teflon cylinder. As shown in Fig. 4, the magnetic field generated from the annular graphene surface sharply decreases to 50 mG, and then slowly decreases, almost zero after about 1000 seconds after turning off the coil power supply.

The strength of the axial magnetic field generated from the annular monolayer graphene wetted with n-hexane sharply decreases, and an axial magnetic field of 50 mG is generated substantially constant from the graphene for the first 30 seconds (see the inset of Fig. 4). Since the strength of the magnetic field in the axial direction is proportional to the magnitude of the circulating current in the annular graphene, this result suggests that the current flowed on the annular graphene surface with zero resistance for 30 seconds. If an annular single layer graphene with an inner diameter of 5 mm and an outer diameter of 10 mm is regarded as a coil with a diameter of 7.5 mm, it can be seen from the constant magnetic flux density (50 mG) in the axial direction that a current of 37.3 mA circulated with no resistance on the annular graphene for 30 seconds. This phenomenon clearly suggests that the superconductivity is obtained by bringing the single-layer graphene into contact with n-hexane. The subsequent decrease in magnetic field does not mean that superconductivity disappeared, but it shows that evaporation of hexane narrowed the region of the superconducting state.

To calculate the self-inductance of the annular single-layer graphene, we assumed that the region where the current flows on the graphene is a rectangular cross-section cylindrical coil with one turn and its height is zero. To calculate the self-inductance of a rectangular cross-section cylindrical coil, it is possible to use an approximate expression of Spielrein [14] that can be applied when the section height is zero. Since the inner diameter and the outer diameter of annular graphene plane are 5 mm and 10 mm, respectively, the self-inductance (L) of the annular graphene is calculated to be $9.487 \times 10^{-9}$ H by using the approximate expression of Spielrein. The time dependence of the current flowing in the circular graphene is obtained by $I(t) = I_0 \exp(-Rt/L)$. Here, $I_0$ is the initial current, R is the resistance, L is the self-inductance, and t is the elapsed time. The sample is a laminated structure composed of p-type silicon having an electric conductivity of 10 S/m, a thermal oxide film $SiO_2$, a single-layer graphene, and n-hexane. Among them, the good electric conductors are graphene and p-type silicon. The conductivity of the single-layer graphene that has been measured is $6.0 \times 10^6$ S/m [15]. The thickness of single-layer graphene can be assumed to be 0.335 nm [16, 17]. Assuming that an annular single-layer graphene having an inner diameter of 5 mm and an outer diameter of 10 mm is replaced approximately by a flat plate having a width of 2.5 mm, a length of 23.56 mm, and a thickness of $3.35 \times 10^{-7}$ mm, since the conductivity of single-layer graphene measured is $6.0 \times 10^6$ S/m, the resistance R of the annular single layer graphene can be approximately regarded as $4.69\times10^3$ Ω. Since the electric conductivity of thermally oxidized silicon ($SiO_2$) is measured as $10^{-14}$ S/m at room temperature [18] and the thickness of the $SiO_2$ film of the sample is 100 nm, the insulation resistance of the $SiO_2$ film is $1.70\times10^{11}$ Ω. Therefore, it can be considered that the single-layer graphene is almost completely insulated from the p-type silicon by the thermal silicon oxide film. Since the resistance R of the annular single-layer graphene can be approximated to $4.69\times10^3$ Ω, so it is found from the time-dependent equation that the magnetic field generated from the annular graphene decays to 0.1% of the initial value at 14.0 ps after cutting off the coil power supply. These facts suggest that the constant circulating current for 30 seconds cannot be explained completely unless it is accepted that superconductivity occurs by bringing n-hexane into contact with the single-layer graphene. The attenuation of the magnetic field generated from the annular graphene does not originate from the disappearance of superconductivity but from the narrowing of the superconducting region due to the evaporation of n-hexane. Therefore, the attenuation of superconducting current by evaporation of n-hexane indicates that the presence of alkane is indispensable for superconductivity. Furthermore, the appearance of superconductivity by binging alkane into contact with graphene suggests that the graphite basal plane is important for superconductivity [19].

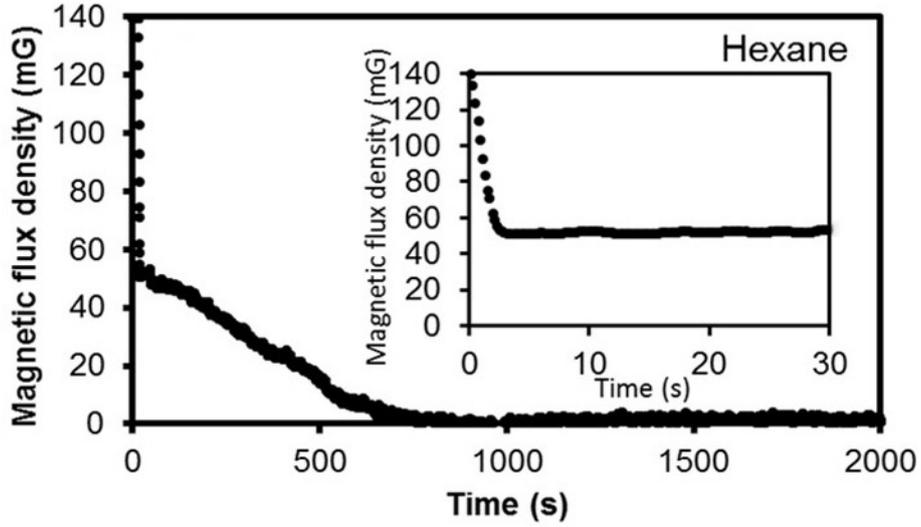

Figure 4 **Changes with time in the magnetic field generated from the annular single-layer graphene that is soaked in n-hexane after turning off the power supply of the coil that applies the magnetic field perpendicular to the graphene surface.** Here, n-hexane was brought into contact with the graphene before the magnetic field was applied to the annular graphene by the coil. The inset shows changes with time of the generated magnetic field from the time of turning off the coil power supply until the subsequent 30 seconds.

## 4. Conclusion

We first applied a magnetic field perpendicular to the single-layer graphene surface at room temperature. Subsequently, we injected n-hexane and n-heptane on the surface of the single-layer graphene. Then we have shown that as soon as the single-layer graphene surfaces are brought into contact with n-hexane or n-heptane, these graphene surfaces exclude the magnetic field completely from the graphene at room temperature. This fact demonstrates that a single-layer graphene brought into contact with alkane shows the Meissner effect at room temperature. The fundamental proof of superconductivity is the demonstration of the Meissner effect. Combining the phenomenon of zero resistance and the phenomenon of sudden change of resistance at a specific temperature which have been shown so far, the demonstration of the Meissner effect clearly proves that graphite brought into contact with alkane is a room temperature superconductor. Furthermore, we wetted the surface of the annular monolayer graphene with hexane, applied a magnetic field to this annular graphene, and then stopped the current flowing through the magnetic field excitation coil. We observed that after the magnetic field excitation coil current was stopped, the magnetic field continued to generate from the annular single-layer graphene and the strength of the magnetic field did not change for the first 30 seconds. This suggests that the single-layer graphene brought into contact with n-hexane exhibits zero resistance at room temperature. After that we observe that the magnetic field gradually decreased. This decrease in the magnetic field is due to the gradual decrease of the superconducting region due to evaporation of n-hexane. It is not due to disappearance of superconductivity in the single-layer graphene bought into contact with n-hexane. The decrease in the magnetic field shows that alkane is indispensable for superconductivity. Superconductivity was observed when alkane was brought into contact with single-layer graphene, indicating that the graphite basal surface is important for superconductivity.